\RequirePackage{silence}
\documentclass[fleqn,usenatbib,useAMS]{mnras} 

\usepackage{newtxtext,newtxmath}

\usepackage{graphicx}
\usepackage{amsmath}

\usepackage{amsfonts}
\usepackage{float}
\usepackage{bm}
\usepackage{ae,aecompl}
\usepackage{array}
\usepackage{soul}
\usepackage{mathtools}
\usepackage{multirow}
\usepackage[T1]{fontenc}
\usepackage[utf8]{inputenc}
\usepackage{booktabs}
\usepackage{graphicx}
\usepackage{subcaption}
\usepackage[normalem]{ulem}

\DeclareRobustCommand{\VAN}[3]{#2}
\let\VANthebibliography\thebibliography
\def\thebibliography{\DeclareRobustCommand{\VAN}[3]{##3}\VANthebibliography}

\DeclareRobustCommand{\appropto}{\mathrel{\vcenter{
		\offinterlineskip\halign{\hfil$##$\cr 
			\propto\cr\noalign{\kern2pt}\sim\cr\noalign{\kern-2pt}}}}}

\defcitealias{Haslbauer_2020}{HBK20}

\title[Joint solution to Hubble \& bulk flow tensions]{A simultaneous solution to the Hubble tension and observed bulk flow within 250~${h^{-1}}$ Mpc} 

\author[S. Mazurenko et al.]{Sergij Mazurenko$^{1}$, Indranil Banik$^{2}$\thanks{E-mail: \href{mailto:sergij.mazurenko@uni-bonn.de}{sergij.mazurenko@uni-bonn.de} (Sergij Mazurenko); \newline $~~~~~~~~~~~~~~~~~\,$ \href{mailto:ib45@st-andrews.ac.uk}{ib45@st-andrews.ac.uk} (Indranil Banik)}, Pavel Kroupa$^{3, 4}$ and Moritz Haslbauer$^{3}$ \vspace{10pt} \\
$^{1}$Universit\"{a}t Bonn, Regina-Pacis-Weg 3, 53115 Bonn, Germany\\
$^{2}$Scottish Universities Physics Alliance, University of Saint Andrews, North Haugh, Saint Andrews, Fife, KY16 9SS, UK\\
$^{3}$Helmholtz-Institut f\"{u}r Strahlen- und Kernphysik, Universit\"{a}t Bonn, Nussallee 14-16, 53115 Bonn, Germany\\
$^{4}$Astronomical Institute, Faculty of Mathematics and Physics, Charles University, V Hole\v{s}ovi\v{c}k\'ach 2, CZ-180 00 Praha 8, Czech Republic}

\date{Accepted XXX. Received YYY; in original form ZZZ}

\pubyear{2023}

\begin{document}
\label{firstpage}
\pagerange{\pageref{firstpage}--\pageref{lastpage}}
\maketitle

\begin{abstract} 
The $\Lambda$ cold dark matter ($\Lambda$CDM) standard cosmological model is in severe tension with several cosmological observations. Foremost is the Hubble tension, which exceeds $5\sigma$ confidence. Galaxy number counts show the Keenan-Barger-Cowie (KBC) supervoid, a significant underdensity out to 300~Mpc that cannot be reconciled with $\Lambda$CDM cosmology. Haslbauer et al. previously showed that a high local Hubble constant arises naturally due to gravitationally driven outflows from the observed KBC supervoid. The main prediction of this model is that peculiar velocities are typically much larger than expected in the $\Lambda$CDM framework. This agrees with the recent discovery by Watkins et al. that galaxies in the CosmicFlows-4 catalogue have significantly faster bulk flows than expected in the $\Lambda$CDM model on scales of $100-250 \, h^{-1}$~Mpc. The rising bulk flow curve is unexpected in standard cosmology, causing $4.8\sigma$ tension at $200 \, h^{-1}$~Mpc. In this work, we determine what the semi-analytic void model of Haslbauer et al. predicts for the bulk flows on these scales. We find qualitative agreement with the observations, especially if our vantage point is chosen to match the observed bulk flow on a scale of $50 \, h^{-1}$~Mpc. This represents a highly non-trivial success of a previously published model that was not constrained by bulk flow measurements, but which was shown to solve the Hubble tension and explain the KBC void consistently with the peculiar velocity of the Local Group. Our results suggest that several cosmological tensions can be simultaneously resolved if structure grows more efficiently than in the $\Lambda$CDM paradigm on scales of tens to hundreds of Mpc.


\end{abstract}


\section{Introduction}

The Hubble tension is one of the greatest currently debated unsolved problems in cosmology \citep{Wong_2020, Migkas_2021, Riess_2022, Valentino_2022}. It is a statistically significant discrepancy between direct local measurements of the Hubble-Lema\^itre constant $H_0$ and the prediction of the Lambda-Cold Dark Matter \citep*[$\Lambda$CDM;][]{Efstathiou_1990, Ostriker_Steinhardt_1995} standard model of cosmology with parameters calibrated to fit the angular power spectrum of anisotropies in the cosmic microwave background (CMB). The local Universe appears to be expanding 10\% faster than this prediction. While the origin of this tension is not known, if the high local determination of $H_0$ is correct, then the universe would have to be about 10\% younger than if the lower CMB-based value is correct. However, the ages of the oldest stars argue against this interpretation \citep{Cimatti_2023}. While their upper limit on $H_0$ is consistent with CMB measurements taken by the Planck satellite \citep{Planck_2020} and the Atacama Cosmology Telescope \citep{Aiola_2020}, it leaves open the issue of why so many local measurements show a faster expansion rate \citep[e.g. figure~1 of][and references therein]{Valentino_2021}. It has recently been argued that the Hubble tension should be solved largely at late times in cosmic history \citep*{Jia_2023, Vagnozzi_2023}, with the expansion rate apparently diverging from $\Lambda$CDM expectations only rather recently \citep{Gomez_2023}.

Another serious but less widely known problem is the Keenan-Barger-Cowie (KBC) void, an underdensity with a diameter of about a Gpc \citep*{Keenan_2013}. Such an extended and deep local underdensity is indicated by several observations at optical \citep{Maddox_1990}, infrared \citep{Huang_1997, Busswell_2004, Frith_2003, Frith_2005, Frith_2006, Keenan_2013, Whitbourn_2014, Wong_2022}, X-ray \citep{Bohringer_2015, Bohringer_2020}, and radio wavelengths \citep*{Rubart_2013, Rubart_2014}. The near-infrared measurements imply that the matter density is only half the cosmic mean value out to a distance of 300~Mpc (see figure~11 of \citealt*{Keenan_2013} and figure~1 of \citealt{Kroupa_2015}). Using data from the Millennium XXL simulation \citep[MXXL;][]{Angulo_2012}, \citet*[][hereafter \citetalias{Haslbauer_2020}]{Haslbauer_2020} showed that the KBC void is in $6.04\sigma$ tension with the $\Lambda$CDM model despite accounting for redshift space distortions induced by outflow from the void, which implies that the actual density contrast is about half that reported by \citet*{Keenan_2013}. Such a deep and extended void suggests a cosmological model in which structure grows faster than in $\Lambda$CDM.

It has been suggested that outflows from a large local void can solve the Hubble tension \citep{Keenan_2016, Shanks_2019, Ding_2020, Camarena_2022, Martin_2023}. This would be a natural solution because a local void creates a hill in the potential, causing galaxies to flow away from the void. From a purely kinematic perspective, the near uniformity of the CMB implies that a substantial underdensity today must be a consequence of outflows. Using this argument, \citetalias{Haslbauer_2020} estimated that the observed KBC void implies that the locally measured $H_0$ should exceed the true background expansion rate by 11\% (see their section~1.1). The Local Group (LG) is situated within the KBC void and is a part of the bulk matter flow. While the Hubble tension can be naturally understood as arising from outflows induced by the KBC void, the problem for $\Lambda$CDM is that it is unable to form such a void because on the relevant scale of 300~Mpc, the universe should be almost homogeneous and isotropic. A simultaneous solution to the KBC void and Hubble tension is possible in $\Lambda$CDM, but only if we assume a $10\sigma$ density fluctuation at early times \citepalias[see figure~1 of][]{Haslbauer_2020}. Moreover, the observations of \citet{Keenan_2013} trace the majority of the galaxy luminosity function, suggesting that the observed underdensity is a genuine underdensity in the total matter distribution (see their figure~9). Allowing for redshift space distortions, \citetalias{Haslbauer_2020} found that the results of \citet{Keenan_2013} imply that the density within 300~Mpc is $\approx 20\%$ below the cosmic mean evident in galaxy number counts at greater distances. This is very much in line with the findings of \citet{Wong_2022}.

\citetalias{Haslbauer_2020} showed that matter inhomogeneities comparable to the KBC void can arise in the neutrino Hot Dark Matter ($\nu$HDM) cosmological framework \citep{Angus_2009, Katz_2013, Wittenburg_2023}. The $\nu$HDM model assumes Milgromian dynamics \citep[MOND;][]{Milgrom_1983, Milgrom_2014_Scholarpedia}. MOND postulates that the gravitational acceleration $g$ in an isolated spherically symmetric system is asymptotically related to the Newtonian gravitational acceleration $g_{_N}$ of the baryons alone according to
\begin{eqnarray}
	g \to \begin{cases}
	g_{_N} \, , & \textrm{if} ~g_{_N} \gg a_{_0} \, , \\
	\sqrt{a_{_0}g_{_N}} \, , & \textrm{if} ~g_{_N} \ll a_{_0} \, .
	\end{cases}
	\label{g_cases}
\end{eqnarray}
The key new ingredient is a fundamental acceleration scale $a_{_0} = 1.2 \times 10^{-10}$~m~s$^{-2}$. This value must be deduced empirically, just like the gravitational constant $G$ in standard gravity. Different studies over the decades have returned very similar values for $a_{_0}$ \citep*{Begeman_1991, Gentile_2011, McGaugh_Lelli_2016}. The MOND gravitational field follows from a Lagrangian, ensuring the usual symmetries and conservation laws with respect to the linear and angular momentum and the energy \citep{Bekenstein_Milgrom_1984, QUMOND}. The least action principle then leads to a generalized Poisson equation that is non-linear in the mass distribution.

MOND has been highly successful on galaxy scales \citep{Famaey_McGaugh_2012, McGaugh_2020, Banik_Zhao_2022}. $\nu$HDM extends its application to cosmology by postulating an extra species of sterile neutrino with a rest energy of 11~eV, which is crucial to matching the high velocity dispersions of virialized galaxy clusters, the offset between the weak lensing and X-ray peaks in the Bullet Cluster, and the high third peak in the angular power spectrum of the CMB \citep[for a review, see section~9.2 of][]{Banik_Zhao_2022}. The background cosmology in $\nu$HDM is the same as in $\Lambda$CDM \citep{Skordis_2019}, though the extra neutrino species implies a mild departure from standard Big Bang nucleosynthesis \citepalias[see section~3.1.2 of][]{Haslbauer_2020}. Moreover, both models behave similarly at early times because when the redshift $z \ga 50$, the typical accelerations are high and thus standard gravity applies. At lower redshifts, significant differences arise in the large-scale structure due to the lack of CDM and the MOND corrections to the gravitational field \citep{Angus_Diaferio_2011, Angus_2013, Katz_2013, Wittenburg_2023}.

Using a semi-analytic model in the $\nu$HDM framework, \citetalias{Haslbauer_2020} evolved three initial void density profiles (Maxwell-Boltzmann, Gaussian, and Exponential) for a large grid of initial void sizes and strengths at $z = 9$. The initial conditions were constrained by observations of the local Universe ($z=0.01-0.15$) and the requirement for the density to almost recover the cosmic mean value at distances of $600-800$~Mpc \citep*[see ﬁgure~11 and table~1 of][]{Keenan_2013}. Any local void solution to the Hubble tension implies quite large peculiar (CMB-frame) velocities, so an important constraint on such models is the observed peculiar velocity of the LG \citep[$v_{\rm{LG}} = 627 \pm 22$~km~s$^{-1}$;][]{Kogut_1993}. While peculiar velocities are typically larger in $\nu$HDM \citep{Katz_2013, Wittenburg_2023}, such a low value arises for a reasonable fraction of observers because gravitational fields from nearby and more distant structures can sometimes partially cancel. However, if we consider a larger region of the universe, the significant Milgromian enhancement to gravity implies much more substantial bulk flows of galaxies on scales of hundreds of Mpc. If this model is to be viable, such rapid bulk flows need to be verified observationally.

We test the local void solution to the Hubble tension proposed in \citetalias{Haslbauer_2020} by extracting the predicted bulk flow from the exact same model without any adjustments in order to compare its \emph{a priori} bulk flow predictions with recently published measurements. Using the CosmicFlows-4 galaxy catalogue \citep{Tully_2023}, \citet{Watkins_2023} present the bulk flow of galaxies on scales of $100-250 \, h^{-1}$~Mpc, where $h \approx 0.7$ is the Hubble constant in units of 100~km~s$^{-1}$~Mpc$^{-1}$, velocities are reported in the CMB frame, and the bulk flow involves a vector average of line of sight peculiar velocities out to some distance (Section~\ref{sec:bulk_flow_velocity}). Bulk flows are expected to decrease with scale, but the observed bulk flow curve has the opposite behaviour and rises to $>400$~km~s$^{-1}$ beyond $160\,h^{-1}$~Mpc. At a scale of $200\,h^{-1}$~Mpc, it is in $4.81\sigma$ tension with the $\Lambda$CDM model ($P = 1.49 \times 10^{-6}$).\footnote{The tension is not stated for a scale of $250 \, h^{-1}$~Mpc, but figure~8 of \citet{Watkins_2023} shows that it exceeds the $5\sigma$ falsification threshold ($P = 5.73 \times 10^{-7}$) typically used in science.} An independent study recently reported ``excellent agreement with the bulk flow measurements of \citet{Watkins_2023}'' using the same dataset and also found significant tension with $\Lambda$CDM at an effective depth of $173 \, h^{-1}$~Mpc, though the more conservative methodology prevented the authors from going further out \citep*{Whitford_2023}. This issue is observationally unrelated to the Hubble tension because adopting a different $H_0$ would affect the peculiar velocities in a spherically symmetric manner, thus not affecting the inferred bulk flow curve shown in figure~7 of \citet{Watkins_2023}. In Section~\ref{sec:results}, we compare the results in the bottom right panel of this figure with the bulk flow predicted by the $\nu$HDM model for different void density profiles and possible vantage points that were previously shown to provide the best overall match to several other cosmological observables \citepalias[mainly the KBC void density profile and the high local $H_0$; see][]{Haslbauer_2020}.

It is important to note that the bulk flow measurements of \citet{Watkins_2023} were not available to \citetalias{Haslbauer_2020}. While earlier bulk flow measurements were available, these were not used $-$ the $\nu$HDM model and the parameters of a local void in this model were in no way constrained to fit the observed bulk flow on scales of several hundred Mpc. The primary objective of this study is to use the bulk flow measurements of \citet{Watkins_2023} to test the \emph{a priori} predictions of the \citetalias{Haslbauer_2020} `Hubble bubble' model on the same scales. This will help to assess whether the Hubble and bulk flow tensions faced by the $\Lambda$CDM model might have a common Milgromian solution. More generally, our results will help to clarify whether the velocity field in a local supervoid solution to the Hubble tension might also match the observed bulk flow on scales of up to $250~h^{-1}$~Mpc.

The structure of this paper is as follows: In Section~\ref{sec:methods}, we describe how we obtain the predicted bulk flows in the same manner as the reported observations. We then present our results in Section~\ref{sec:results} and conclude in Section~\ref{sec:conclusions}.

\section{Methods}
\label{sec:methods}

The starting point for our analysis is the peculiar velocity field of the semi-analytic void model of \citetalias{Haslbauer_2020}, which is shown in their figure~8 for the Maxwell-Boltzmann void density profile (velocity fields of the relevant models are shown in our Appendix~\ref{sec:velocity_fields}). This is a combination of a spherically symmetric outflow from the void centre with a systemic motion of the whole void towards the left, which reduces the spherical symmetry to axisymmetry. We refer to the peculiar velocity in the CMB frame as the total velocity $\bm{v}_{\rm{tot}}$ \citepalias[analogous to equation~72 of][]{Haslbauer_2020}. This is well suited for a comparison with observations because in studies of the bulk flow on large scales, heliocentric redshifts are typically corrected for the precisely known velocity of the Sun in the frame of the CMB.\footnote{R. Watkins, private communication.}

\subsection{Possible locations for the Local Group}
\label{sec:possible_LG_locations}

The position of the LG in this framework can be deduced by finding points which move with a velocity of $v_{\rm{tot}} = v_{\rm{LG}} = 627 \pm 22$~km~s$^{-1}$ \citep{Kogut_1993}, where we use the notation that $v \equiv \left| \bm{v} \right|$ for any vector $\bm{v}$. Figure~8 of \citetalias{Haslbauer_2020} shows the locus of such points with a solid black curve. Those authors suggested that the LG is towards the right hand side of this curve, which is located further away from the void centre. However, their study lacked any reliable way to precisely determine where the LG ought to be.

We simplify our analysis by only considering two possible LG locations, which we call the inner and outer vantage point (relative to the void centre). These bracket the range of possible LG distances from the void centre. Importantly, both vantage points place the LG on the symmetry axis of the problem, allowing us to work with an axisymmetric code. We consider both vantage points separately for all three considered void (under)density profiles (Maxwell-Boltzmann, Gaussian, and Exponential). We also consider uncertainties in the model predictions for each of these six cases due to the uncertainty in $v_{\rm{LG}}$, which can slightly vary the location of the LG by a few Mpc (see Appendix~\ref{sec:velocity_fields}). This is however less significant than observational uncertainties in the bulk flow.

\begin{table}
    \centering
    \caption{The distance of the LG from the void centre (in Mpc) in each of the considered void density profiles for the two possible vantage points consistent with the CMB-frame velocity of the LG (see the text). The impact of its 22~km~s$^{-1}$ uncertainty \citep{Kogut_1993} is considered separately in all cases. We also consider setting $v_{\rm{LG}} = 840$~km~s$^{-1}$ (Section~\ref{Vantage_point_discussion}), in which case the inner vantage point's distance from the void centre is 115.99 (158.39)~Mpc for the Gaussian (Exponential) void profile.}
    \label{tab:possible locations of LG for density profiles}
    \begin{tabular}{llccc}
        \hline
        & & \multicolumn{3}{c}{Density profile} \\
        LG location & $v_{\rm{LG}}$ (km/s) & MB & Gauss & Exp \\ \hline
        Inner & $627 + 22$ & 136.38 & 142.59 & 192.19 \\
        vantage & 627 & 138.08 & 145.79 & 196.39 \\
        point & $627 - 22$ & 139.78 & 149.29 & 200.69 \\ \hline
        Outer & $627 + 22$ & 265.37 & 505.95 & 848.56 \\
        vantage & 627 & 262.27 & 494.76 & 816.16 \\
        point & $627 - 22$ & 259.17 & 483.86 & 786.46 \\ \hline
    \end{tabular}
\end{table}

In Table~\ref{tab:possible locations of LG for density profiles}, we show the possible locations of the LG for all three void density profiles. In all cases, the inner vantage point lies within 200~Mpc of the void centre. The outer vantage point is generally much more distant. Only the Maxwell-Boltzmann profile gives an outer vantage point reasonably close to the void centre, at a distance of $d_{\rm{MB},r} = 262.27$~Mpc.

\subsection{Observed peculiar velocity}
\label{sec:observed_vpec}

We describe the problem using Cartesian coordinates centred on the observer. Only two axes are needed because the problem is axisymmetric. The $x$-axis corresponds to the symmetry axis of the problem, while the $y$-axis lies in the orthogonal direction. The distance of any point from the observer is thus $r = \sqrt{x^2 + y^2}$.

Observations of distant galaxies can only tell us their peculiar velocity along the line of sight. Applying this consideration, the observable component of the peculiar velocity at point $i$ is
\begin{eqnarray}
    v_{\rm{obs}, i} ~=~ \left( \frac{xv_x + yv_y}{r} \right)_i \, ,
    \label{eq:observable velocity component}
\end{eqnarray}
where $\bm{v}_{\rm{tot}} \equiv \left( v_x, v_y \right)$ is the velocity in the CMB frame. This is found by adding the velocity relative to the void centre with the systemic velocity of the void as a whole, which is towards $-x$. The systemic void velocity is given in tables~4 and C1 of \citetalias{Haslbauer_2020} and is not adjustable in our analysis.

\subsection{Bulk flow velocity}
\label{sec:bulk_flow_velocity}

While the line of sight peculiar velocity of point $i$ is typically thought of as a scalar quantity, it will help to think of it as a vector $\bm{v}_{\rm{obs}, i}$ pointing from the observer towards point $i$. The bulk flow velocity in the CMB frame is the weighted average of these line of sight peculiar velocity vectors within a spherical volume of radius $r_{\rm{bulk}}$ centred on the observer. This definition matches equation~26 of \citet{Watkins_2023}.

Due to the simulated velocity field being axisymmetric with respect to the $x$-axis and the observer also lying on this axis, the bulk flow must be along it. We therefore consider only the $x$-component of $\bm{v}_{\rm{obs}, i}$, which is given by
\begin{eqnarray}
    v_{{\rm obs,x},i} ~=~ v_{{\rm obs},i} \left( \frac{x}{r} \right)_i \, .
    \label{eq:observable velocity x-component}
\end{eqnarray}
Calculating the bulk flow velocity is then just a matter of taking a suitably weighted average of this quantity, which we discuss below.

Each point has a weight $w_i \propto V_i/r^2$, where $V_i$ is the volume covered by cell $i$. The factor of $1/r^2$ is necessary because we are dealing with the radial velocity components instead of the 3D velocities. Under the assumption that the velocity field is curl-free, this will then match the average of the 3D velocities, which is how the bulk flow is defined \citep*[see section~3 of][]{Peery_2018}. To minimize numerical issues at very low radii caused by the $1/r^2$ factor, we soften it by adding to $r$ in quadrature half the radial width of each cell used in our calculation. Due to the high resolution used, this has a negligible impact on our analysis. The bulk flow is then
\begin{eqnarray}
    v_{\rm{bulk}} \left( \leq r_{\rm{bulk}} \right) ~=~ \frac{\sum_{i} w_i v_{\rm{obs},x,i}}{\sum_{i} w_i} \, ,
    \label{eq:bulk flow velocity}
\end{eqnarray}
where the sums are taken over all points within $r_{\rm{bulk}}$ of the observer. The volume weighting accounts for the inhomogeneous distribution of points, a consequence of the discretization scheme detailed in section~3.3.6 of \citetalias{Haslbauer_2020} and reused here for simplicity. This scheme reduces the computational cost by working with a polar coordinate system centred on the void centre. Because we need to work with spherical regions centred on the LG rather than the void centre, we substantially increase the resolution compared to that used in \citetalias{Haslbauer_2020}. We also compute the velocity field out to a larger distance because we need predictions out to $250\,h^{-1}$~Mpc from the outer vantage point.

Since the simulated velocity field is smooth, setting $r_{\rm{bulk}} \to 0$ implies that all considered points have the same $\bm{v}_{\rm{tot}}$. This velocity must be parallel to the symmetry axis as the velocity component perpendicular to it must become negligible at small distances from the axis. Equations~\ref{eq:observable velocity component} and \ref{eq:observable velocity x-component} can then be combined and simplified to give
\begin{eqnarray}
    v_{\rm{obs},x,i} ~=~ v \left( \frac{x}{r} \right)^2_i ~=~ v \cos^2{\alpha_i} \, ,
    \label{observable_velocity_for_small_radii}
\end{eqnarray}
where $v \equiv \left| \bm{v}_{\rm{tot}} \right|$ is the speed of any particle in the small region under consideration, $\alpha_i$ is the angle at the observer between the direction towards point $i$ and the positive $x$-axis, and the power of 2 accounts for the projection of $\bm{v}_{\rm{tot}}$ onto the line of sight and the projection of the line of sight onto the direction of $\bm{v}_{\rm{tot}}$. Relating this to Equation~\ref{eq:bulk flow velocity} shows that the observed bulk flow in this small region is only $v/3$ because the angle-averaged value of $\cos^2{\alpha}$ is 1/3 when there are 3 spatial directions. Note also that in a small sphere centred on the observer, we must have that $\bm{v}_{\rm{tot}} \to \bm{v}_{\rm{LG}}$ as $r_{\rm{bulk}} \to 0$ because we expect that the peculiar velocity of the LG arises mostly from structure on quite large scales. In the model, $\bm{v}_{\rm{LG}}$ is towards $-x$ for the inner vantage point and towards $+x$ for the outer vantage point. This is because the systemic velocity of the void dominates near its centre, while outflow from the void becomes more important further out.

\section{Results and Discussion}
\label{sec:results}

\begin{figure*}
    \includegraphics[width=\textwidth]{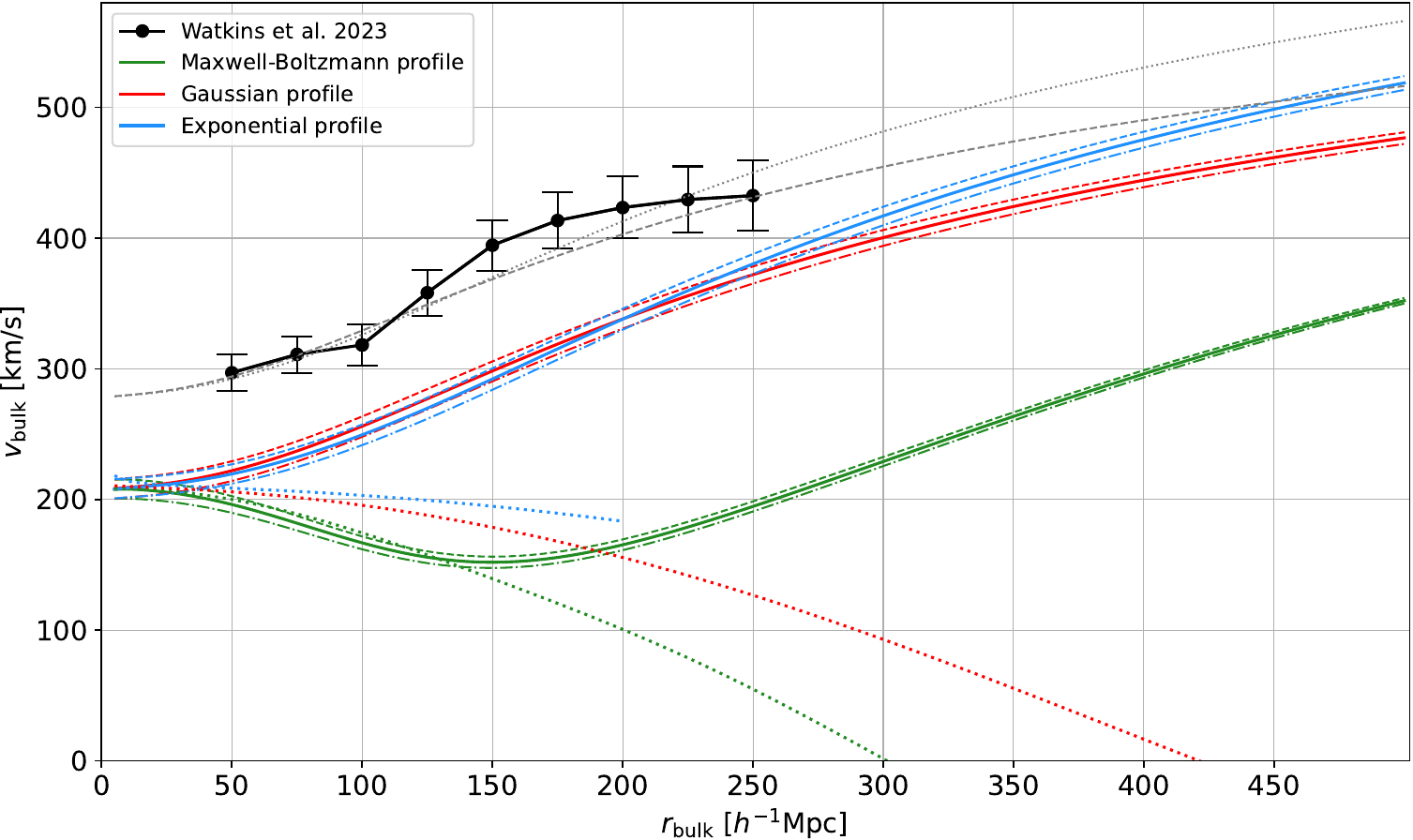}
    \caption{Bulk flows in spheres of different radii around an observer with a suitable CMB-frame velocity for the Maxwell-Boltzmann, Gaussian, and Exponential density profiles (green, red, and blue curves, respectively). The solid black points with uncertainties show the observed bulk flows \citep[bottom right panel of figure~7 in][]{Watkins_2023}. The two innermost data points use the same method and survey data but were not published in that study, so these were provided in a private communication by its lead author. At $r_{\rm{bulk}} = 0$, the simulated curves begin at $v_{\rm{LG}}/3$ because the observed bulk flows use only the line of sight velocities, which are treated as vectors that are then combined (see below Equation~\ref{observable_velocity_for_small_radii}). The solid curves represent the bulk flows as seen by the inner observer if $v_{\rm{LG}} = 627$~km~s$^{-1}$. The dashed curves (typically above each solid curve) are the bulk flows for $v_{\rm{LG}} = 627 + 22$\,km\,s$^{-1}$. The dashed-dotted curves (typically below each solid curve) are the bulk flows for $v_{\rm{LG}} = 627 - 22$\,km\,s$^{-1}$. These additional curves help to show the impact of the uncertainty in $v_{\rm{LG}}$, which slightly shifts our vantage point in the context of each model and can thus affect the predicted bulk flow. The grey dashed (dotted) curve shows the bulk flow curve for the Gaussian (Exponential) density profile if $v_{\rm{LG}} = 840$~km~s$^{-1}$, which shifts the inner vantage point by $29.8 \left( 38.0 \right)$~Mpc towards the void centre. The coloured thin dotted curves represent the outer observers for all density profiles. These all assume the observed $v_{\rm{LG}}$ and use the same colour as the corresponding results for the inner observer.}
    \label{fig:Comparison_Watkins2023-AllObservers-AllProfiles}
\end{figure*}

The bulk flow curves for all considered void profiles and LG locations (inner/outer) are shown in Figure~\ref{fig:Comparison_Watkins2023-AllObservers-AllProfiles} for the nominal $v_{\rm{LG}}$. Results for the Maxwell-Boltzmann, Gaussian, and Exponential void profile are shown in green, red, and blue, respectively. Results for the inner (outer) vantage point are shown using a solid (dotted) line. Two additional curves are shown for the inner vantage point in which $v_{\rm{LG}}$ is varied from its nominal value by its observational uncertainty: the dashed-dotted (dashed) curves assume a lower (higher) $v_{\rm{LG}}$. The resulting uncertainty on the predicted bulk flow is much smaller than in the observed bulk flow amplitude \citep[solid black points with error bars;][]{Watkins_2023}. All the simulated bulk flow curves start at $v_{\rm{LG}}/3$ for reasons discussed below Equation~\ref{observable_velocity_for_small_radii}.

The bulk flow represents an average of the motions within a region. If the velocity field were linear in the position, then the bulk flow curves would remain flat at $v_{\rm{LG}}/3$. Non-linearity in the velocity field causes the bulk flow curve to deviate from this value. To better understand the simulated bulk flow curves, it will help to consider the velocity field (Appendix~\ref{sec:velocity_fields}).

We first consider the case where the LG is at the inner vantage point. The bulk flow curve for the Maxwell-Boltzmann profile is fairly flat because $v_{\rm{tot}}$ is nearly linear: if we consider its value along the symmetry axis for simplicity, it drops from $\approx 1500$~km~s$^{-1}$ at the void centre to zero at $\approx 200$~Mpc (as indicated by the black star) and then rises back up again at a similar rate, albeit with a velocity in the opposite direction (Figure~\ref{fig:velocity_map_MB_profile}). This causes the average motion within a sphere to stay roughly flat as we increase the radius of the sphere, leading to a fairly flat bulk flow curve. Since the observed bulk flows are much faster than $v_{\rm{LG}}/3$ and increase with scale, it is not possible for the Maxwell-Boltzmann profile to explain the observations (solid green curve in Figure~\ref{fig:Comparison_Watkins2023-AllObservers-AllProfiles}).

The situation is rather different for the Gaussian and Exponential profiles, which yield bulk flow curves that rise rather quickly. If we again consider $v_{\rm{tot}}$ along the symmetry axis, it rapidly drops from $\approx 2000$~km~s$^{-1}$ at the void centre to 627~km~s$^{-1}$ at the location of the LG, but $v_{\rm{tot}}$ then only very gradually decreases to zero (see Figures~\ref{fig:velocity_map_GAUSS_profile} and \ref{fig:velocity_map_EXP_profile}, respectively). This causes the average motion within a sphere to rise quickly with the radius of the sphere, leading to a rising bulk flow curve. We expect the bulk flow to remain parallel to $\bm{v}_{\rm{LG}}$ at all radii, which is approximately the case \citep{Kogut_1993, Watkins_2023}.

We find that both the Gaussian and Exponential density profiles result in bulk flow curves with a similar trend to the observed bulk flow curve: there is an initial steep increase which then almost stagnates at large radii. However, both models undershoot the observations by $50-100$~km~s$^{-1}$, with the largest deviation at $r_{\rm{bulk}} \approx 150\,h^{-1}$~Mpc. Part of this can be attributed to relativistic corrections, which imply that the actual bulk flows are about 10\% smaller than reported \citep{Heinesen_2023}. However, these corrections scale with $z$ and are in any case too small to fully explain the discrepancy.

\subsection{Our vantage point}
\label{Vantage_point_discussion}

The bulk flow around the outer vantage point completely differs from the inner one, as shown by the continuously declining thin dotted curves in Figure~\ref{fig:Comparison_Watkins2023-AllObservers-AllProfiles} (the same colour is used for each density profile). If the LG were located at the outer vantage point, then regardless of the void density profile, the bulk flow would decrease with $r_{\rm{bulk}}$ and reverse direction at $r_{\rm{bulk}} \approx 300 \, h^{-1}$~Mpc ($420 \, h^{-1}$~Mpc) for the Maxwell-Boltzmann (Gaussian) density profile. The bulk flow curves for the outer vantage point are in catastrophic disagreement with the observed fact that $v_{\rm{bulk}} \gg v_{\rm{LG}}/3$ for $r_{\rm{bulk}} \approx 200 \, h^{-1}$~Mpc. The outer vantage points are at least $2\times$ further away from the void centre than their respective inner counterparts ($4\times$ for the Gaussian and Exponential profiles). Thus, our results seem to imply that we need to be within $\approx 200$~Mpc of the void centre (see Table~\ref{tab:possible locations of LG for density profiles}), in agreement with the conclusion of \citetalias{Haslbauer_2020}. Being fairly close to the centre of a void would also limit the extent to which the Universe appears anisotropic.

The impact of the 22~km~s$^{-1}$ uncertainty in $v_{\rm{LG}}$ is illustrated in Figure~\ref{fig:Comparison_Watkins2023-AllObservers-AllProfiles}, where the dashed (dashed-dotted) lines show the impact of raising (reducing) $v_{\rm{LG}}$ by this amount from its nominal value of 627~km~s$^{-1}$. The uncertainty in $v_{\rm{LG}}$ has little impact on the observed bulk flow for the inner observers because the modified velocity of the LG only slightly shifts its position (see Table~\ref{tab:possible locations of LG for density profiles} and Appendix~\ref{sec:velocity_fields}). The resulting changes to $v_{\rm{bulk}}$ are $<10$~km~s$^{-1}$. Increasing $v_{\rm{LG}}$ by its uncertainty causes the bulk flow curves for the Gaussian and Exponential profiles to slightly shift towards the observational data, while decreasing $v_{\rm{LG}}$ has the opposite effect. However, these differences are barely perceptible. 


The bulk flow curves from the Gaussian and Exponential density profiles underestimate the observed bulk flow curve. Note that since the bulk flow measurements are already in the CMB frame, the assumed $v_{\rm{LG}}$ enters our analysis only by setting our location within the void. The closer the vantage point is to the centre of the void, the higher the bulk flow curve becomes overall. Considering the inner vantage point with any density profile, the uncertainty in $v_{\rm{LG}}$ of 22~km~s$^{-1}$ corresponds to a shift to our position by a mere $3 - 4$~Mpc. As discussed earlier, this is barely perceptible.

If we allow a larger positional shift of the LG, we could in principle match the observed bulk flow curve fairly well. Such an increase to $v_{\rm{LG}}$ is plausible because ideally our location within the void would be set using an estimate of $v_{\rm{local}}$, the bulk flow on a scale large enough to average over motions at the scale of individual galaxy groups and clusters but small compared to the KBC void. We assumed $v_{\rm{local}} = v_{\rm{LG}}$ as an approximation due to the LG frame being the largest reference frame to which a transformation can easily be performed. However, the true value of $v_{\rm{local}}$ will generally differ from $v_{\rm{LG}}$. The investigated model by design ignores the possibility of small-scale structures as it only considers the much larger structure that is the KBC void. In the real Universe, additional structures like gravitational attractors exist in the form of galaxy clusters, for instance the Virgo and Fornax clusters at distances of 17~Mpc and 20~Mpc, respectively \citep{Mei_2007, Blakeslee_2009}. Small-scale underdensities can also play a role. In the local vicinity of such small-scale structures, the velocity field would become distorted, thereby shifting the appropriate vantage point. The larger scale velocity field of the void would remain unaffected by this.

Interestingly, the bulk flow on a scale of $50\,h^{-1}$~Mpc is observed to be $297 \pm 14$~km~s$^{-1}$ \citep[private communication from R. Watkins; same method and data as in][]{Watkins_2023}. This suggests an actual bulk motion on this scale of $891 \pm 42$~km~s$^{-1}$ based on our argument in Equation~\ref{observable_velocity_for_small_radii}. Allowing for the fact that the predicted bulk flow curve rises by just under 15~km\,s$^{-1}$ between $r_{\rm{bulk}} = 0$ and $r_{\rm{bulk}} = 50\,h^{-1}$~Mpc for the most realistic void models from \citetalias{Haslbauer_2020}, the above result suggests that an appropriate choice of $v_{\rm{local}}$ is slightly below 891~km~s$^{-1}$. This motivates us to consider the case $v_{\rm{LG}} = 840$~km~s$^{-1}$ and update our vantage point accordingly. The results are shown using the dashed (dotted) grey line in Figure~\ref{fig:Comparison_Watkins2023-AllObservers-AllProfiles} for the Gaussian (Exponential) profile. In both cases, the model agrees quite well with the observed bulk flow curve, perhaps indicating that the local velocity due to large-scale structure is more accurately reflected by the bulk flow on a $50\,h^{-1}$~Mpc scale than by $\bm{v}_{\rm{LG}}$. Regardless of the considered void models, the mismatch between the observed $v_{\rm{LG}}$ and $3 v_{\rm{bulk}}$ suggests that the LG is moving $\approx 200$~km~s$^{-1}$ slower in the CMB frame than is typical for the matter within $50\,h^{-1}$~Mpc. This modest difference could plausibly arise from nearby structures beyond the LG.

\subsection{Predictions at large radii}
\label{Distant_predictions}

For all considered density profiles, the \citetalias{Haslbauer_2020} model predicts a rising bulk flow amplitude over the range $r_{\rm{bulk}} = 250 - 500 \, h^{-1}$~Mpc if we are located at the inner vantage point, as strongly suggested by our results. At $500 \, h^{-1}$~Mpc, the Gaussian and Exponential profiles reach bulk flow amplitudes of approximately 475~km~s$^{-1}$ and 520~km~s$^{-1}$, respectively. However, these predictions should be treated with caution because it is likely that our simplified model for the KBC void breaks down at such large distances as there must be other structures in the Universe. Even within the model we are using, it is necessary to assume more distant structures that impose an external gravitational field on the void as a whole \citepalias{Haslbauer_2020}. At some point, it is no longer possible to approximate the structures responsible for this external field as being much more distant than the region under consideration. These issues should be more thoroughly addressed using a Gpc-scale cosmological simulation of the $\nu$HDM model and analysis of KBC void analogues therein (Russell et al., in preparation). It will be interesting to see if the bulk flow typically reverses on larger scales than those considered by \citet{Watkins_2023}, as would be needed to explain the dipole anisotropy in the galaxy cluster distance-redshift relation \citep{Migkas_2021}.

A local void solution to the Hubble tension necessarily implies that the Planck cosmology works at suitably high redshift. While our focus here is on the predictions of the \citetalias{Haslbauer_2020} model at the present epoch, those authors used the lightcone analysis described in their section~3.3.3 to show that the void-induced enhancement to the redshift decays rather slowly and is not negligible 8~Gyr ago (see their figure~16). This corresponds to $z \approx 1$, which can seem far beyond the reach of a local void. But if the void is fairly large and given also the rather slower decline of MOND gravity with distance compared to Newtonian gravity, it is quite possible that the impact of the void on cosmological observables is not restricted to $z < 0.1$, as was assumed in figure~1 of \citet*{Kenworthy_2019}. It is thus not surprising that those authors were unable to find any evidence of a local void \citepalias[other problems with a local void scenario were addressed in detail in section~5.3 of][]{Haslbauer_2020}. It is interesting in this regard that the data of \citet*{Kenworthy_2019} only goes up to $z = 0.5$ (see their figure~5). More recent studies do find a change to the inferred $H_0$ at higher redshift if the data are binned in $z$ and care is taken to infer the cosmological parameters from only the data within each redshift bin, removing residual correlations with the data in other bins \citep*{Jia_2023}. Their analysis clearly shows a return to a Planck cosmology at higher redshift, regardless of whether we consider supernovae alone or combine it with other datasets to improve the accuracy. A similar result is apparent in the preliminary analysis of \citet{Gomez_2023}, whose reconstructed expansion rate history closely follows a Planck cosmology at high redshift and deviates only when $z \la 0.6$. We are currently exploring the consistency of the \citetalias{Haslbauer_2020} model with cosmological datasets at high redshift, building on its known success fitting supernova data out to $z = 0.15$ (Mazurenko et al., in preparation).

\section{Conclusions}
\label{sec:conclusions}

It has been suggested that the anomalously high local $H_0$ could be due to outflows from a local void \citepalias{Haslbauer_2020}. We revisit their best-fitting semi-analytic void model for the three initial ($z = 9$) density profiles they considered (Maxwell-Boltzmann, Gaussian, and Exponential). In each case, we consider two possible vantage points where the simulated peculiar velocity in the CMB frame matches that of the LG. For these six combinations of void density profile and LG location, we determine the bulk flow $v_{\rm{bulk}}$ in spheres of different radii $r_{\rm{bulk}} \leq 250 \, h^{-1}$~Mpc. This predicted bulk flow curve is compared with recent observations \citep{Watkins_2023} based on the CosmicFlows-4 galaxy catalogue \citep{Tully_2023}, which is designed to obtain reliable distances and thus line of sight peculiar velocities. We also consider the small uncertainty in the predicted bulk flow due to the uncertainty in $v_{\rm{LG}}$, which slightly shifts the location of the LG. The impact on the predicted $v_{\rm{bulk}}$ is much smaller than its observational uncertainty (Figure~\ref{fig:Comparison_Watkins2023-AllObservers-AllProfiles}).

With all three density profiles, the predicted bulk flow curve from the outer vantage point (relative to the void centre) is strongly incompatible with observations, both qualitatively and quantitatively. Assuming instead that we are located at the inner vantage point, the Maxwell-Boltzmann density profile also gives a bulk flow curve that is observationally excluded (albeit in somewhat better agreement). However, the Gaussian and Exponential profiles give bulk flow curves that qualitatively agree quite well with the observational data, nicely capturing the steep increase followed by a flattening out. Within the context of the \citetalias{Haslbauer_2020} model, this implies a void which is deepest in the centre and that we are located within $200$~Mpc of the void centre while remaining consistent with the observed LG peculiar velocity. Taking into account its 22~km~s$^{-1}$ uncertainty barely impacts the bulk flow curve. However, if one allows for small-scale flows within the void which are not captured by the model (as discussed in Section~\ref{Vantage_point_discussion}), then a good quantitative agreement is achievable (grey curves in Figure~\ref{fig:Comparison_Watkins2023-AllObservers-AllProfiles}). These curves nicely match the bulk flow measurement of \citet{Watkins_2023} on a scale of $50 \, h^{-1}$~Mpc, suggesting that the velocity of the LG has been reduced slightly by local structures on a scale not considered in the model.


The agreement outlined above is reasonable considering the simplicity of the semi-analytic model used, which assumes spherical symmetry relative to the void centre plus a constant extra velocity everywhere within the void due to the impact of even more distant structures \citepalias{Haslbauer_2020}. Obviously the local Universe does not have such a simplified velocity field, which must instead be tied to its complex history of structure formation over a Hubble time. This should be explored using a Gpc-scale cosmological simulation of the $\nu$HDM paradigm, building on previously published smaller scale simulations \citep[e.g.,][]{Katz_2013, Wittenburg_2023} $-$ but ideally reaching scales similar to the MXXL simulation in the $\Lambda$CDM framework \citep{Angulo_2012}. This will address the plausibility of the observed bulk flows in a self-consistent $\nu$HDM cosmological simulation (Russell et al., in preparation). In addition, detailed future observations of the bulk flow beyond $250 \, h^{-1}$~Mpc can be compared to the bulk flows predicted by the here utilized models and those found in Gpc-scale $\nu$HDM simulations. Once such measurements become available, the tension with the $\Lambda$CDM model should also be quantified. It is quite possible that future data will worsen the $4.8\sigma$ tension reported by \citet{Watkins_2023} on a scale of $200 \, h^{-1}$~Mpc, especially given that the bulk flow seems to be increasing with scale while the typically expected value in $\Lambda$CDM decreases with scale (see their figure~7).

It is important to stress that the recent bulk flow measurements of \citet{Watkins_2023} were not available to \citetalias{Haslbauer_2020}, who did not in any way constrain the void parameters using the observed bulk flow. Despite this, their model naturally accounts for the observed rising behaviour of the bulk flow on scales of $100 - 250 \, h^{-1}$~Mpc without any parameter adjustments for two of the six considered combinations of void profile and observer location, which was chosen to match the peculiar velocity of the LG. This agreement suggests that a Milgromian cosmological framework can consistently explain the Hubble and bulk flow tensions by assuming that we are located in a large supervoid similar to the observed KBC void. Our analysis also imposes constraints on the position of the LG relative to the KBC void centre, and to a lesser extent on the void density profile. Future work will test this model through better measurements of the KBC void density profile and the bulk flows on larger scales than those probed so far. In addition, it will be important to explore self-consistent cosmological simulations rather than the simplified semi-analytic model considered here.

Our results suggest that many of the severe cosmological tensions faced by $\Lambda$CDM could be alleviated if structure grows more rapidly than it predicts. This could also explain the unexpectedly large integrated Sachs-Wolfe effect associated with supervoids \citep{Kovacs_2019, Kovacs_2022_ISW} and the 2\% dipole in the distribution of quasars \citep{Secrest_2021, Secrest_2022, Dam_2023}. In this scenario, overdensities should also be more pronounced than expected in standard cosmology. There is strong evidence for this from massive high-redshift galaxy cluster collisions like El Gordo, which has been shown to falsify $\Lambda$CDM at $>5\sigma$ confidence for any plausible infall velocity \citep*{Asencio_2021, Asencio_2023}. Even the formation of individual galaxies could be faster than expected in $\Lambda$CDM, as suggested by several recent discoveries with the JWST of massive galaxies in the very early Universe \citep[$z \ga 10$;][and references therein]{Haslbauer_2022_JWST}. Such rapid galaxy formation can be naturally understood in MOND \citep{Eappen_2022}. While it is too early to say what might be responsible for these tensions with $\Lambda$CDM, the wide range of different observational techniques used to reveal these inconsistencies makes it likely that many if not all of them are genuine problems for the paradigm that go beyond the background expansion history.

\section*{Acknowledgements}

IB is supported by Science and Technology Facilities Council grant ST/V000861/1. PK thanks the Deutscher Akademischer Austauschdienst-Eastern European exchange programme for support. The authors thank Alfie Russell and Harry Desmond for helpful discussions. They are also very grateful to the referee Richard Watkins for comments and suggestions that significantly improved this paper.


\section*{Data Availability}

The simulation results were previously published in \citetalias{Haslbauer_2020}. The observed bulk flows and their uncertainties were provided by Richard Watkins based on \citet{Watkins_2023}.



\bibliographystyle{mnras}
\bibliography{Bulk_flow_bbl}


\begin{appendix}

\section{Void velocity fields}
\label{sec:velocity_fields}

\begin{figure}
    \includegraphics[width=\columnwidth]{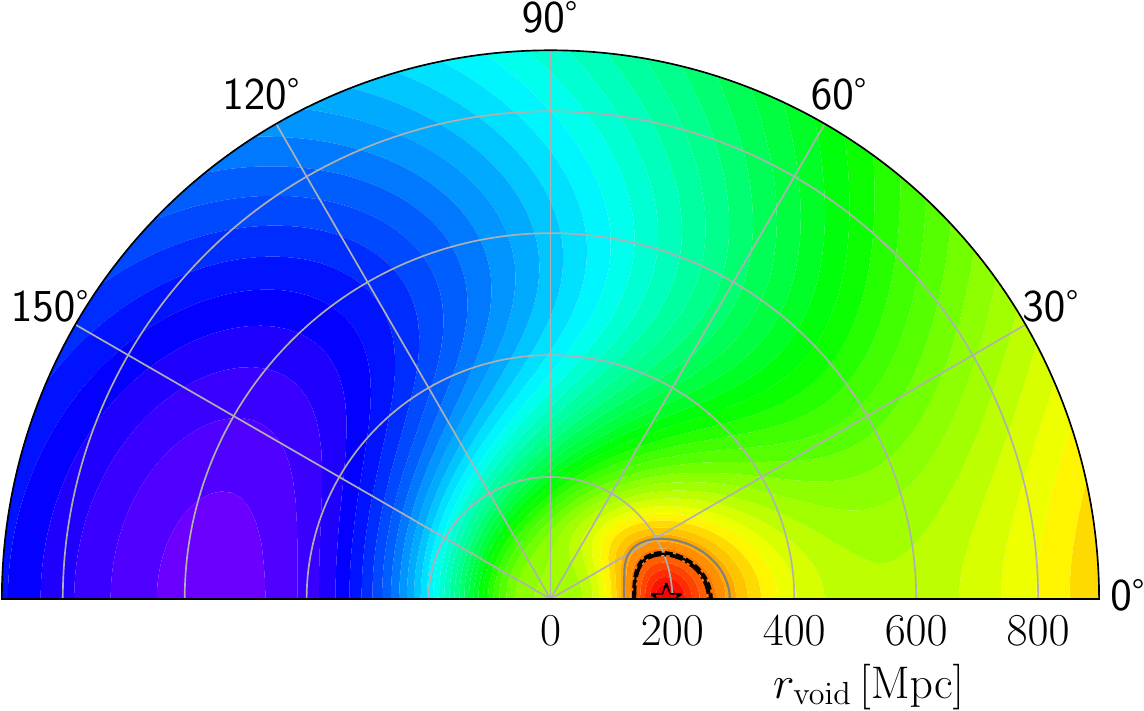}
    \includegraphics[width=\columnwidth]{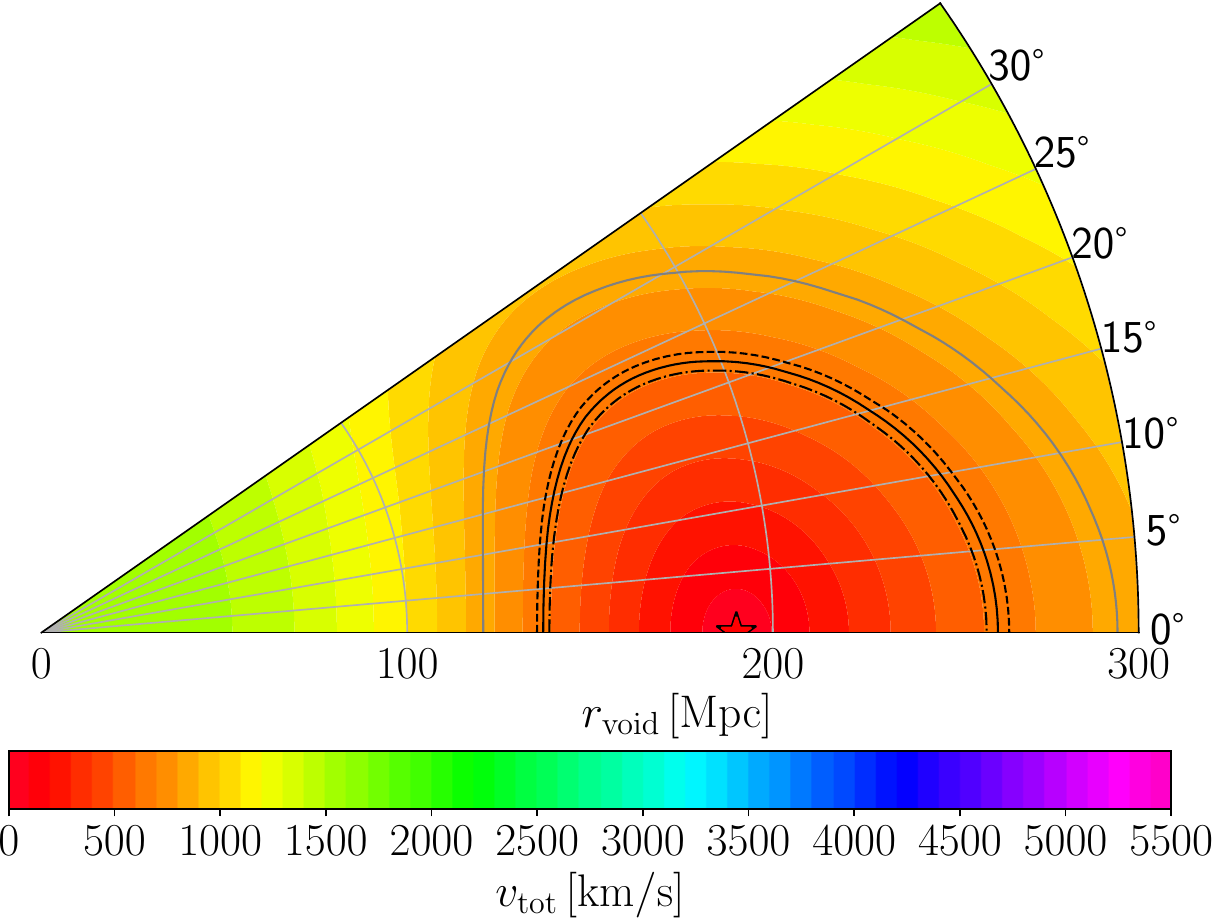}
    \caption{Map of the total velocity $v_{\rm tot}$ in the CMB frame resulting from the Maxwell-Boltzmann void density profile. Results are axisymmetric about the $x$-axis because the model applies a Galilean transformation to a spherically symmetric outflow \citepalias{Haslbauer_2020}. Distances and angles shown here are relative to the void centre. The solid black curve shows the locus of points where $v_{\rm tot} = v_{\rm{LG}}$. To highlight its uncertainty, the dashed-dotted (dashed) line shows the locus of points where $v_{\rm{LG}}$ is assumed to be smaller (larger) than its observed value by its observational uncertainty. The grey line corresponds to $v_{\rm{LG}} = 840$~km~s$^{-1}$ (Section~\ref{Vantage_point_discussion}). The bottom panel shows a zoomed in view. $\bm{v}_{\rm tot}$ is directed roughly away from the black star in both panels marking the point where $v_{\rm tot} = 0$.}
    \label{fig:velocity_map_MB_profile}
\end{figure}

\begin{figure}
    \includegraphics[width=\columnwidth]{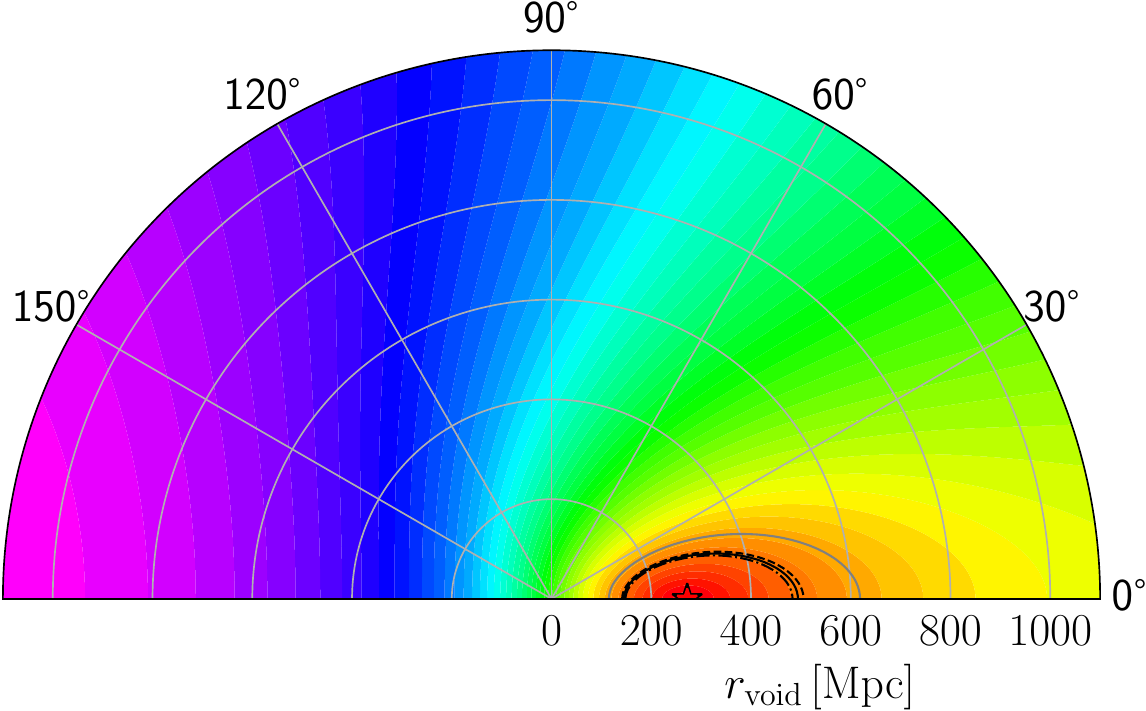}
    \includegraphics[width=\columnwidth]{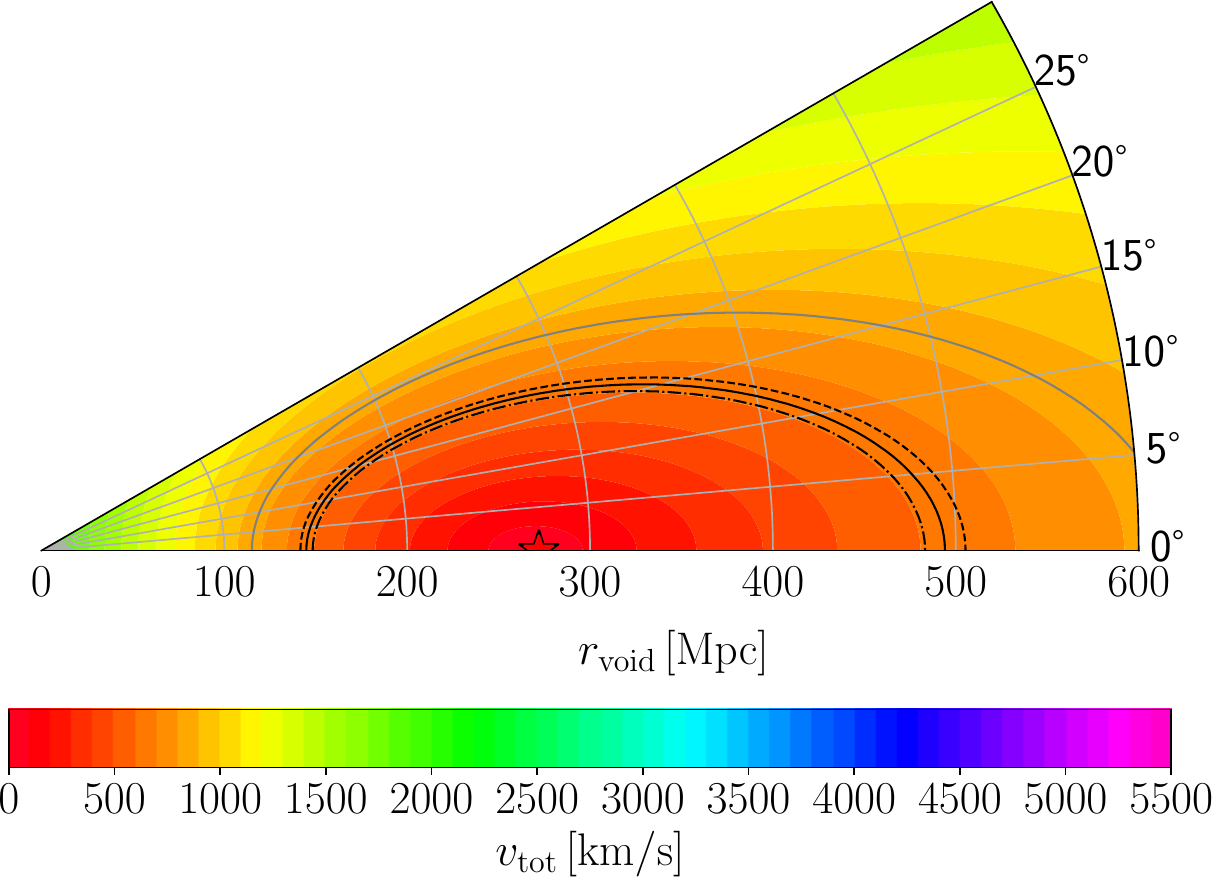}
    \caption{Similar to Figure~\ref{fig:velocity_map_MB_profile}, but for the Gaussian density profile. Notice that the isovelocity contours are elongated towards the right, indicating that the velocity field changes more slowly at larger distances from the void centre. This causes the rising bulk flow curve evident in Figure~\ref{fig:Comparison_Watkins2023-AllObservers-AllProfiles} (see Section~\ref{sec:results}).}
    \label{fig:velocity_map_GAUSS_profile}
\end{figure}

\begin{figure}
    \includegraphics[width=\columnwidth]{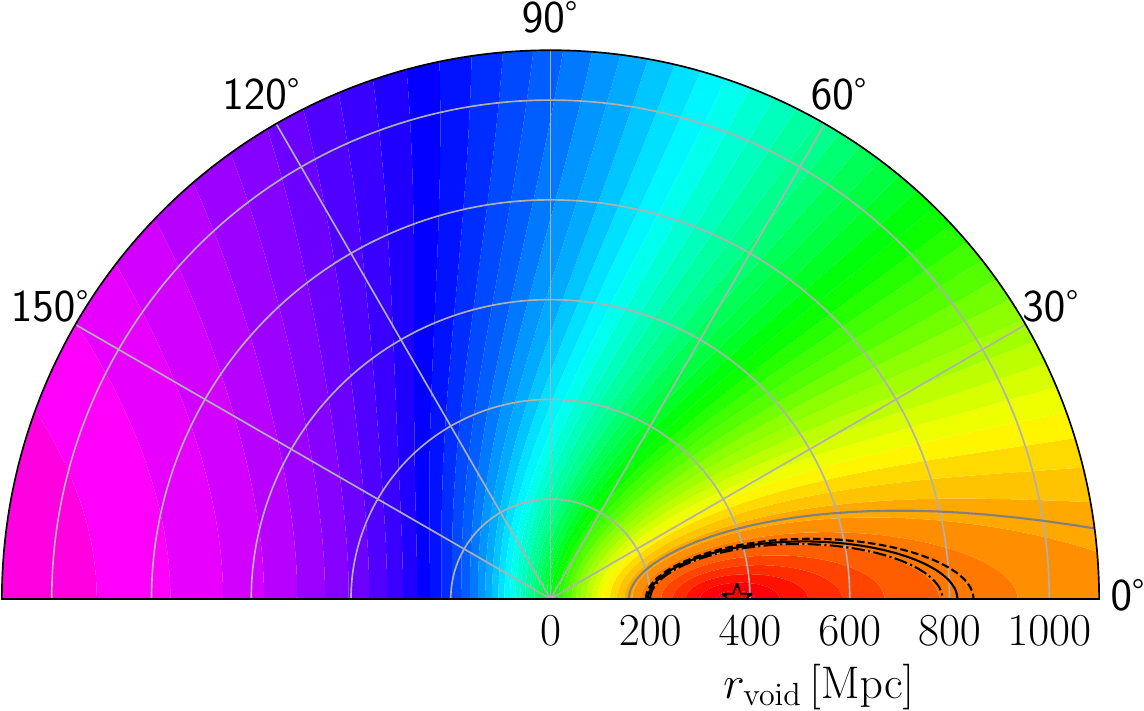}
    \includegraphics[width=\columnwidth]{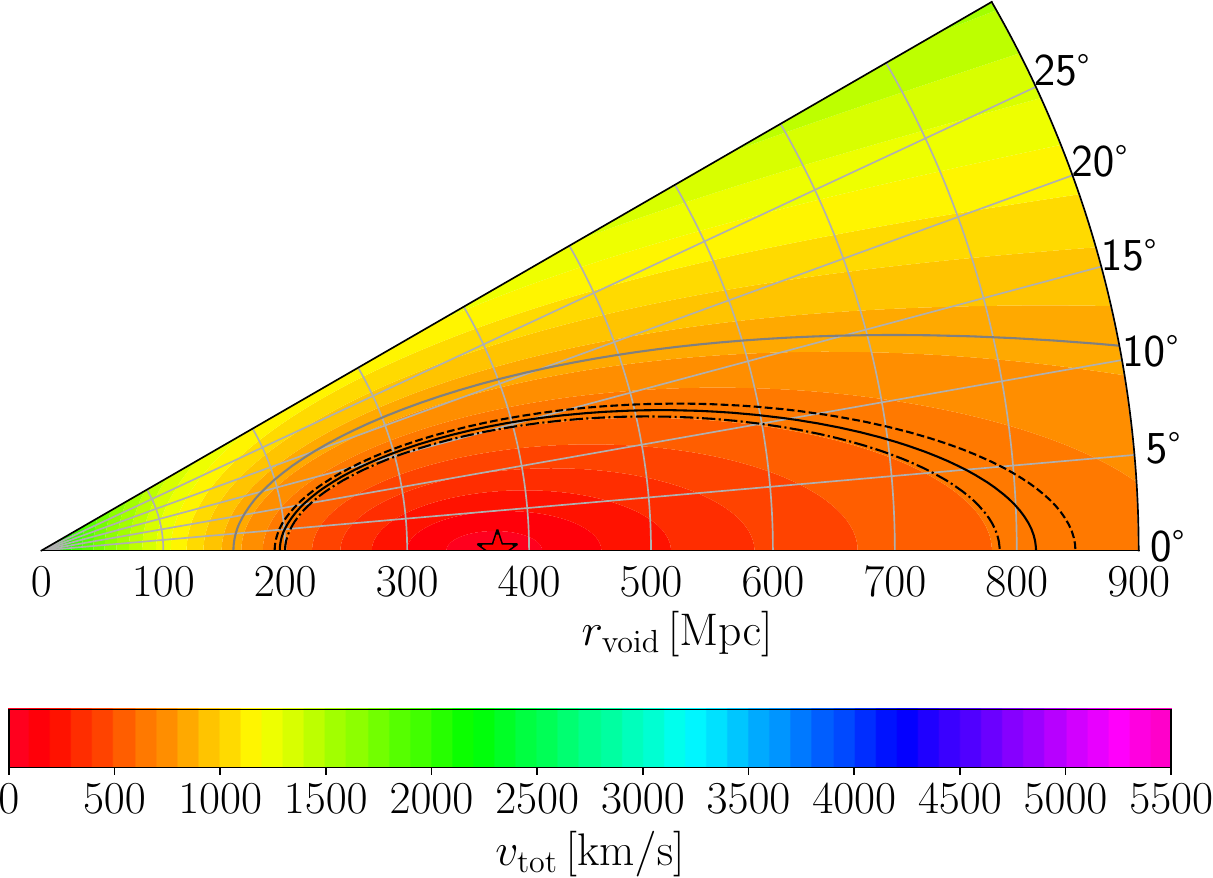}
    \caption{Similar to Figure~\ref{fig:velocity_map_MB_profile}, but for the Exponential density profile.}
    \label{fig:velocity_map_EXP_profile}
\end{figure}

While our focus has been on the bulk flow, it can be helpful to consider the velocity field in the void models we explore. We therefore show a 2D map of $v_{\rm{tot}}$ for the Maxwell-Boltzmann, Gaussian, and Exponential density profiles in Figures~\ref{fig:velocity_map_MB_profile}, \ref{fig:velocity_map_GAUSS_profile}, and \ref{fig:velocity_map_EXP_profile}, respectively. The velocity fields are axisymmetric around the $x$-axis, with the void as a whole moving towards the left. The velocity fields are shown in polar coordinates centred on the void centre, which is situated at $r_{\rm{void}} = 0$. While the direction of $\bm{v}_{\rm{tot}}$ is not shown for simplicity, this is directed approximately radially away from the point where $v_{\rm{tot}} = 0$, which we indicate with a black star. In all these figures, the solid black line shows the locus of points where $v_{\rm{tot}} = v_{\rm{LG}} = 627$~km~s$^{-1}$, while the dashed and dashed-dotted black lines represent the curves along which $v_{\rm{ tot}} = 627 + 22$~km~s$^{-1}$ and $v_{\rm{tot}} = 627 - 22$~km~s$^{-1}$, respectively, thus bracketing the observational uncertainty in $v_{\rm{LG}}$. The grey line represents $v_{\rm{LG}} = 840$~km~s$^{-1}$, which is motivated by the bulk flow measurement of \citet{Watkins_2023} on a scale of $50 \, h^{-1}$~Mpc (see Section~\ref{Vantage_point_discussion}). The lower panels show a close-up of the region where $v_{\rm tot} = 627 \pm 22$~km~s$^{-1}$. The vantage points considered for the bulk flow calculations are at the two intersections of each isovelocity contour with the $x$-axis. Our results strongly suggest that the LG is within 200~Mpc of the void centre.

\end{appendix}

\bsp 
\label{lastpage}
\end{document}